# Measurement of the second-order coherence function for metallic nanolasers

WILLIAM E. HAYENGA[1], HIPOLITO GARCIA-GRACIA[1], HOSSEIN HODAEI[1], CHRISTIAN REIMER[2], ROBERTO MORANDOTTI[2], PATRICK LIKAMWA[1], MERCEDEH KHAJAVIKHAN[1*]

[1] *CREOL, The College of Optics and Photonics, University of Central Florida, Orlando, FL 32816-2700, USA.*
[2] *INRS-EMT, 1650 Boulevard Lionel Boulet, Varennes, Québec J3X 1S2, Canada*

*Corresponding author: mercedeh@creol.ucf.edu*

**Due to the high spontaneous emission coupled into the resonance mode in metallic nanolasers, there has been a debate concerning the coherence properties of this family of light sources. The second-order coherence function can unambiguously determine the nature of a given radiation. In this paper, an approach to measure the second-order coherence function for broad linewidth sources in the near-infrared telecommunication band is established based on a modified Hanbury Brown and Twiss configuration. Using this set-up, it is shown that metallic coaxial and disk-shaped nanolasers with InGaAsP multiple quantum well gain systems are indeed capable of generating coherent radiation. © 2016 Optical Society of America**

## 1. INTRODUCTION

In recent years, there has been tremendous progress towards the development of metallic and metallo-dielectric nanoscale lasers [1-16]. These advances are largely motivated by their small footprint and potential for high-speed operation, which makes such nanolasers great candidates for on-chip sources in photonic integrated circuits [16]. Through the use of metal as cladding, the volume of the laser cavity can be reduced to subwavelength dimensions without significantly compromising the mode confinement ($\Gamma$). If designed properly, the portion of the spontaneous emission coupled into the lasing mode ($\beta$) can even approach unity, in which case the laser is known to be "thresholdless" [11].

Determining the onset of coherent emission in such high-$\beta$ resonators can prove challenging [17,18]. Typically, a light-light (L-L) or light-current (L-I) curve, where the output power is measured as a function of incident light or injection current, can be used to resolve whether a light-emitting device is a laser. In general, when the L-L or L-I curve is plotted in a logarithmic scale, it will exhibit an "S" shape, which consists of three regions. At the lower left end is the photo-luminescence (PL) dominant region where the L-L or L-I curve can be represented by a line having a theoretical slope of one, in the middle is a sharp jump in the intensity due to the prevalence of amplified spontaneous emission (ASE), and at the upper right end is the lasing region where the curve regains its unity slope. However, these three regimes are only readily discernible in lasers having a low spontaneous emission coupling factor ($\beta \ll 1$). As $\beta$ increases, the sharpness due to the ASE begins to soften. When $\beta \rightarrow 1$, the nonlinearity from the ASE completely disappears and the ensuing curve becomes a line in its entirety [11, 19, 20]. Consequently, merely assessing the L-L or L-I curve is no longer a viable approach for determining the lasing threshold. It should be noted that the situation depicted above, with unity slopes in the PL and lasing regions, is only valid under the assumption that most processes involved are radiative and no thermal roll-over is present. In the majority of semiconductor lasers, where the non-radiative recombination processes cannot be neglected, it is expected that the slope of the lines to deviate from unity at the lower end of the PL (due to surface recombination), and at the higher end of the lasing regime (due to Auger recombination) – a set of trends that can make the L-L or L-I curve of a light-emitting diode (LED) to appear like that of a laser.

An unambiguous measure to determine the nature of a given emission is the second-order coherence function ($g^2(\tau) = \langle I(t)I(t+\tau)\rangle/\langle I(t)\rangle^2$), which is an intensity correlation function of the radiation. The intensity fluctuations of light are in general classified as chaotic ($g^2(0) > 1$), coherent ($g^2(0) = 1$), or sub-Poissonian ($g^2(0) < 1$) [21]. In this regard, the second-order coherence differs fundamentally from the first-order coherence function which serves as a description of phase. For example, a white thermal source spectrally filtered to such a degree that the temporal coherence ($\tau_c$) is equal to that of a laser may be deemed classically coherent. However, still present in this emission would be the amplitude fluctuations associated with the statistical nature of the photons' arrival. The second-order coherence function can be used to further characterize the emission properties of light – in particular lasers. For a laser device, below threshold, light is expected to be super-Poissonian (bunched), while near the classically defined threshold the emission transitions to a coherent state *i.e.* the photons' arrival follow a Poissonian distribution [18].

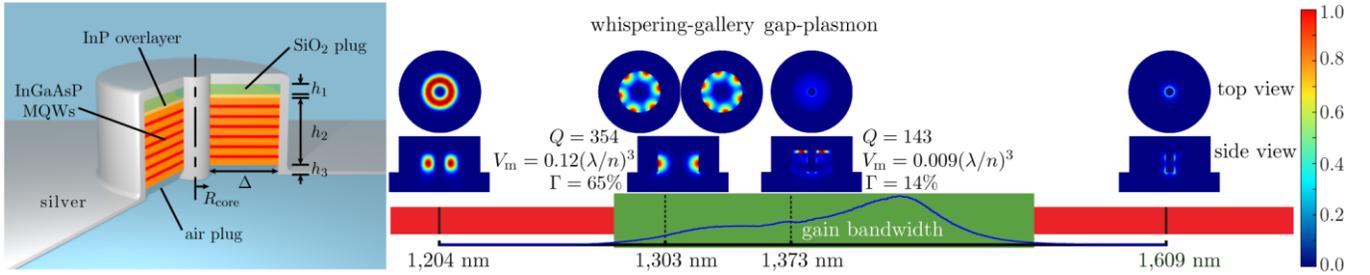

**Fig. 1.** Coaxial nanolaser geometry and modes. **a,** Illustration and **b,** modal content of the metallic coaxial nanolaser. The resonator supports three modes within the gain bandwidth of the active medium: a pair of degenerate modes at 1303 nm and a gap-plasmon-type mode at 1373 nm. Q and $\Gamma$ denote the quality factor and the extent of energy confinement to the semiconductor region, $V_m$, the effective modal volume. The color bar shows normalized $|E|^2$, where $E$ is the electric field. Nominal permittivity values are used in this simulation.

So far, the second-order coherence function has been measured for a number of dielectric-based nanolasers. In particular, the $g^2(\tau)$ has been obtained for photonic crystal and micropillar cavities either by using standard Hanbury Brown-Twiss (HBT) interferometry [22-24], through second-harmonic generation [25], or even by recording the arrival of photons using a streak camera [26]. In all above cases, the gain system is composed of quantum dots that are capable of generating emissions of narrow linewidth at cryogenic temperatures. It should be noted that a narrow linewidth is crucial for such second-order coherence measurements since the defining characteristics of the function only appears within the coherence time of the emitted light. In addition, at the same output power level, narrow linewidth entails larger spectral density. The $g^2(\tau)$ function has also been measured for a spaser-type device composed of an InGaN/GaN nanorod on a silver substrate, operating at visible wavelengths [12]. In general, what makes second-order coherence measurements more challenging for metallic nanolasers is their broad emission linewidth and low output power. This problem becomes yet more acute for nanolasers operating in the near-infrared regime where both the efficiency and the timing resolution of the detectors are considerably lower than their visible counterparts.

In this paper, the second-order coherence properties of metallic coaxial and disk-shaped nanolasers are explored near and above their classically defined lasing threshold. Section 2 describes the coaxial laser cavity design, and its first-order optical properties. In Section 3 an approach for measuring the second-order coherence of broad linewidth radiation sources, using the HBT method, is established. This setup is then used in Section 4 to determine the nature of the emitted light from a number of multiple quantum-well metal-clad nanolasers including the laser characterized in Section 2. Section 5 concludes the paper.

## 2. COAXIAL NANOLASER: DESIGN AND CHARACTERIZATION

So far, lasing operation has been demonstrated in a number of metal-coated nanocavities [1,2,5,7,9-11]. However, there has been a debate whether the emitted light from these structures is truly coherent [27]. Generally, such viewpoints are motivated by the relatively broad linewidth of the emission, and the lack of readily distinguishable regions in the L-L curve. This is particularly the case in metallic coaxial nanolasers that can simultaneously exhibit a high $\beta$, low quality factor (Q-factor), and high $\Gamma$ [11].

Figure 1a shows a schematic of the coaxial nanolaser under study. It is comprised of a metallic rod ($R_{core}$: 50 nm) surrounded by a metal-coated semiconductor ring ($\Delta$: 200 nm, $h_2$: 210 nm). The gain-medium (ring) consists of six vertically stacked quantum wells with an overall height of 200 nm, each composed of a 10 nm thick well (In$_{x=0.56}$Ga$_{1-x}$As$_{y=0.938}$P$_{1-y}$) sandwiched between two 20 nm thick barrier layers (In$_{x=0.734}$Ga$_{1-x}$As$_{y=0.57}$P$_{1-y}$). The quantum wells are covered by a 10 nm thick InP overlayer for protection. The upper and lower ends of the ring are terminated by silicon dioxide (SiO$_2$) and air plugs (heights $h_3$: 30 nm and $h_1$: 20 nm, respectively). The PL spectrum of the bare quantum well system is measured at several pump powers at a temperature of 77 K. The PL spectrum is depicted in the spectral window of gain in Fig. 1b at a pump power comparable to that required to reach lasing operation in our devices. The modal content of the nanolaser is obtained using electromagnetic (EM) simulations based on the finite element method (FEM) and is displayed in Fig. 1b. The EM simulations are performed with material parameters at a temperature of 77 K (permittivity of silver: $-90 - 1.0i$, quantum well gain system: 11.35, InP: 9.8, and SiO2: 2.2). Our simulations indicate that the coaxial resonator under study supports three modes within the gain bandwidth of the active medium: two degenerate whispering-gallery-type modes at 1303 nm as well as a gap-plasmon-like mode at 1373 nm. It should be noted that small variations of permittivities (<2%) and dimensions (<5%) around the above nominal values do not change the modal content within the gain bandwidth (only their corresponding wavelengths and to a small extent their Q-factors are affected). In this simulation, $\Gamma$ is determined through an appropriate normalized overlap integral, and the effective modal volume is calculated as $V_m = \int_{V_a} dv \varepsilon(r)|E(r)|^2 / \max\{\varepsilon(r)|E(r)|^2\}$, where $V_a$ is the volume of the active region [28]. Although all three modes can potentially participate in the lasing process, the higher Q-factor, as well as the larger $\Gamma$ of the set of the degenerate modes places them first in line to reach the lasing threshold. The spontaneous emission coupling factor for this cavity is found by calculating the emission from a randomly oriented dipole in a random location within the active region. $\beta$ is estimated as the ratio of the emitted power at the wavelength of the desired lasing mode to the total power radiated by the dipole, weighted by the photoluminescence profile of the bare quantum well system. For the modes at 1303 nm, the calculated $\beta$ is divided by half to account for the degeneracy. This procedure is repeated for ten random dipoles and the results are averaged to find a $\beta \sim 0.048$. It should be noted that the above value of $\beta$ is still considerably larger than most micro-scale semiconductor lasers, e. g. for VCSELs $\beta < 10^{-3}$. While coaxial structures with somehow different dimensions can exhibit higher spontaneous emission coupling factors, in this work, we focused our attention to the above device because of its higher output power due to its improved outcoupling. The higher output power allows us to study the second-order coherence properties closer to the threshold condition.

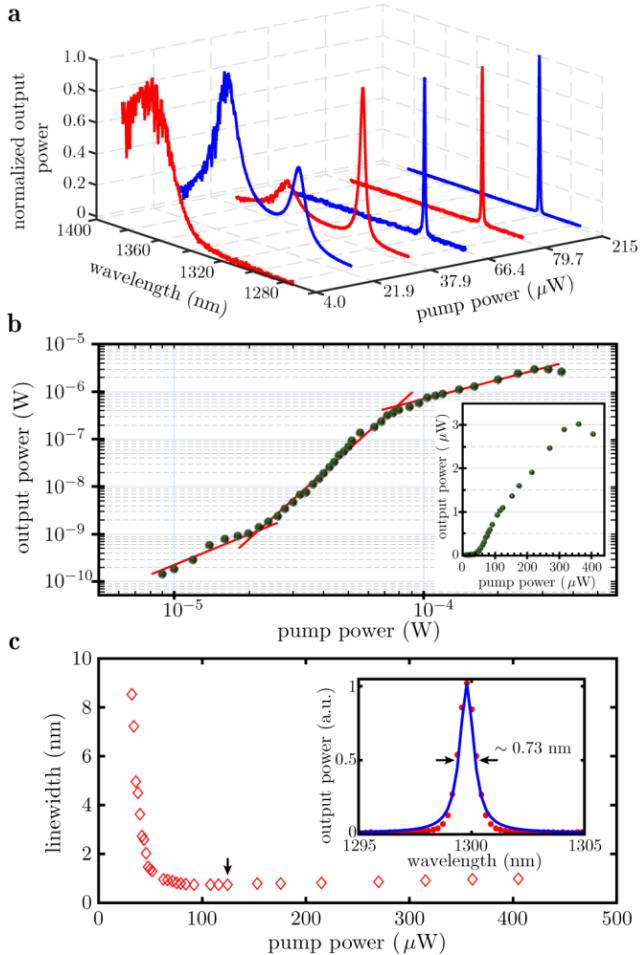

**Fig. 2.** Characterization of the nanolaser under CW optical pumping at a temperature of 77 K. **a,** the spectral evolution of the laser, **b,** the light-light curve in a logarithmic scale and linear scale (inset), as well as **c,** the linewidth versus pump power. The Lorentzian fit used to estimate the linewidth is depicted in the inset of **c**.

The coaxial laser is fabricated using the method outlined in [11], and characterized in a micro-photoluminescence (μ-PL) set-up to collect the evolution of the spectrum (Fig. 2a), the L-L curve (Fig. 2b), and the linewidth (Fig. 2c). The nanolasers are pumped optically using a continuous wave (CW) single-mode fiber laser operating at 1064 nm. All nanolasers reported in this manuscript are cooled to a temperature of 77 K – mainly to boost the laser efficiency in order to be able to perform the subsequent second-order coherence measurements. The spectral evolution of the above laser is shown in Fig. 2a. At lower pump powers the gap-plasmon mode at 1373 nm is the first resonance to appear in the PL, because spectrally it is closer to the peak of the gain. However, due to its smaller Q-factor and lower $\Gamma$, this mode does not reach the threshold condition. As the pump power increases, one of the modes at 1303 nm emerges – ultimately dominating the other modes. This behavior of the modes is in excellent agreement with the simulation results of Fig. 1b. The L-L curve of this laser is plotted in a logarithmic scale in Fig. 2b and linear scale in the inset. The three characteristic regions of the logarithmic L-L curve are emphasized with lines plotted in Fig. 2b – the PL near the lower left corner, the ASE in the center, and the lasing in the upper right. The upsurge in output power that is expected in the ASE region begins around 20 μW and softens at 70 μW, where the device appears to transition into lasing operation. The emission power from the laser continues to increase, until ultimately Auger recombination and to some extent thermal roll-over become predominant, causing the output power to decrease. It should be noted that the output power reported in Fig. 2b presents the actual power collected off the sample via an objective with a numerical aperture (NA) of 0.42 and intensity transmission of ~57% at ~1300 nm. Due to the limited NA of the objective lens, it is estimated that the power at the exit aperture of the laser is about ten times greater than the values provided in Fig. 2b. Finally, the measured linewidth of the emitted light is plotted in Fig. 2c, showing a sharp decrease in the emission linewidth until it levels off at around 80 μW. The inset of Fig. 2c displays the Lorentzian fit used in determining the linewidth. Far above threshold, as the pump power increases, the linewidth slightly broadens. This behavior is almost universal in all the nanolasers we studied, both coaxial and disk-shaped, and to some extent may be attributed to the optical pumping scheme that introduces heating and carrier fluctuations. For the above reported laser, the minimum measured linewidth is ~ 0.7 nm. This relatively broad linewidth is a byproduct of the large $\Gamma$ and high $\beta$. In this regard, a large portion of the spontaneous emission lands in the lasing mode. This noise is then amplified by the gain due to the high mode confinement. In fact, as suggested in a new study, an effective strategy to achieve an ultra-narrow linewidth in semiconductor lasers is through reducing both $\Gamma$ and $\beta$ [29].

## 3. SECOND-ORDER COHERENCE MEASUREMENT SET-UP AND CALIBRATION

To further investigate the nature of the radiation from nanolasers, a modified Hanbury Brown-Twiss set-up is prepared. In this set-up, the light under study is split and guided into two arms of the interferometer where each arm is equipped with a single photon avalanche diode (SPAD). Upon the arrival of a photon, the first SPAD triggers a time-correlated single photon counting (TCSPC) module operating in start-stop mode, setting the start point. When the second SPAD detects a photon, a stop signal is sent to the TCSPC module, and the resulting time delay ($\tau$) shows the arrival correlation of the photons. For chaotic light, at zero time delay ($\tau = 0$) a coincidence peak is expected – a direct result of photon bunching. On the other hand, for coherent radiation with a Poissonian distribution, the resulting correlation function is expected to be unity for all time delays. It should be noted that, similar to other interferometric set-ups, the aforementioned coherence properties can only be observed within the coherence time of the emission. Consequently, in order to measure the second-order coherence of broad linewidth sources, either the detection system must have very good timing resolution (in the order a few tens of femtoseconds), or the emission must first be spectrally filtered such that its temporal coherence becomes larger than the timing resolution of the SPADs and TSCPC module [30]. Limited by the timing resolution of the currently available single photon counters, we chose to spectrally filter the radiation to extend the coherence time of the radiation under study. Clearly, spectral filtering when performed out of the laser cavity cannot alter the statistical distribution of photons' arrival [31].

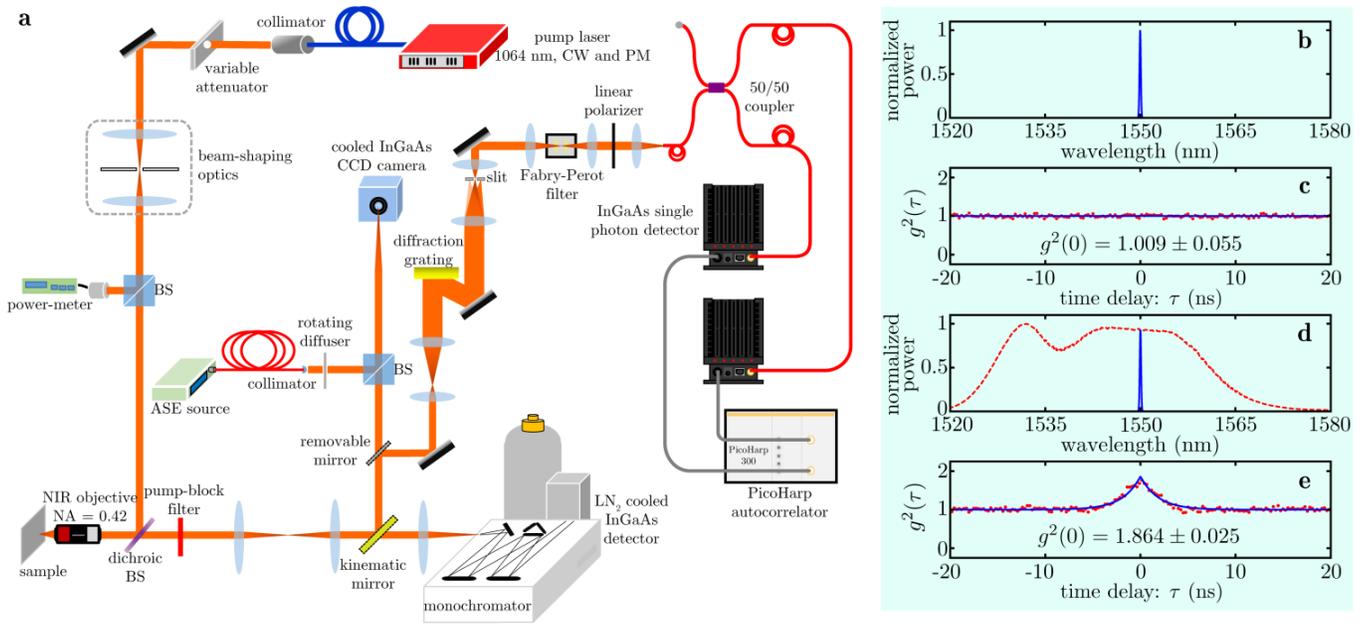

**Fig. 3.** Measurement setup and calibration results. **a,** the μ-PL set-up with the incorporated Hanbury Brown-Twiss interferometer. **b,** Schematic of the spectrum and **c,** second-order coherence measurement for the Agilent 81460A laser. **d,** Measured output spectrum (dashed) and representation of the filtered emission (solid), and **e,** second-order coherence function, for the ASE source.

The experimental set-up, which includes both a μ-PL characterization station and a modified HBT interferometer, is illustrated in Fig. 3a. The HBT interferometer is incorporated into the μ-PL set-up via a kinematic mirror that redirects the light through a spectral filtering stage. The filter is comprised of a cascaded arrangement of a diffraction grating and a Fabry-Perot (Thorlabs SA210-12B) to increase the temporal coherence to ~ 3.75 ns, well beyond the resolution of the SPADs (IDQ ID220, resolution: 240 ps). A linear polarizer is used to remove undesired polarization components. The spectrally filtered light is then collected using a single-mode fiber, split into equal parts with a 50:50 directional coupler, and directed to the SPADs through fibers each equipped with a variable optical attenuator (VOA). Lastly, the SPADs are connected to a TCSPC module (PicoHarp 300) to collect the time-correlated histograms. The overall loss in the set-up, associated with the HBT interferometer, is found to be ~ 69 dB – this includes both the losses related to the optical components after the kinematic mirror (~ 36 dB) as well as the power filtered out due to the linewidth narrowing (~ 33 dB, ultimately depending on the spectral linewidth of the radiation). A detailed description of the losses in the system can be found in Supplementary Information Part 1.

In order to establish the second-order coherence capability of the set-up, we first measure the $g^2(\tau)$ for a commercially available laser (Agilent 81460A). The laser is set to generate an output emission at 1550 nm with a linewidth of 50 MHz (using the coherence control module) – schematically shown in Fig. 3b. In this case the measured $g^2(\tau)$ function is a flat line across all time delays – confirming that the source is indeed a laser (Fig. 3c). We then measure the second order coherence function for a commercially available ASE source (Amonics ALS-CL-20). The output from the ASE source has a spectral width of ~50 nm centered at ~1550 nm as displayed in Fig. 3d – yielding a coherence time on the order of 100 fs. Without spectral filtering, the $g^2(\tau)$ of the ASE source also resulted in a flat line (featureless) – falsely resembling that of a coherent source. However, by using the filtering scheme incorporated in the set-up, the emitted radiation from the ASE source is narrowed down to ~85 MHz. For this spectrally narrowed emission the second-order coherence function is no longer a line. Instead, it clearly shows a coincidence peak of $g^2(0) = 1.864 \pm 0.025$ (Fig. 3e). In this case, the collected coincidence data (presented with dots in the figure) are fitted using the Siegert relation $g^2(\tau) = 1 + |g^1(\tau)|^2$, where $|g^1(\tau)|^2 = \exp(-2|\tau|/\tau_c)$ [30]. From this fitting, the temporal coherence was estimated to be $\tau_c = 3.75$ ns – agreeing well with the linewidth expected from the cascaded diffraction grating and the Fabry-Perot filter. The fact that a broadband ASE source with intrinsic spectral linewidth of ~50 nm is capable of demonstrating a $g^2(0) \sim 1.864$ confirms that the current setup can be used for characterizing the second-order coherence properties of an arbitrary source with a broad linewidth.

## 4. SECOND-ORDER COHERENCE MEASUREMENT FOR METAL-CLAD NANOLASERS

After calibrating the HBT set-up to reliably characterize the statistical nature of the light from an arbitrary source with broad linewidth, the second-order coherence measurements are performed for a number of coaxial and disk-shape lasers with various radii. It should be noted that in order to be able to compare the results and to ensure the accuracy of the measurements, most of the intensity correlation data are collected over the same period of time (~7 minutes) and count rate (~50 KHz).

Fig. 4 shows the measured second order coherence function at different pump levels, along with the logarithmic single-shot emission spectrum, for the coaxial nanocavity described in section 2. Far above threshold at a pump power of 215 μW, the measured $g^2(\tau)$ is a flat line, where the fit suggests a $g^2(0) = 1.009 \pm 0.038$ (Fig. 4a) – confirming that the coaxial structure under study, to a good approximation, is capable of generating coherent radiation. A similar measurement at a pump power of 79.7 μW shows a slightly increased $g^2(0)$ of $1.037 \pm 0.039$ (Fig. 4c). Finally, the second-order coherence measured barely below the classically defined threshold, at a pump power of 66.4 μW, yields $g^2(0) = 1.081 \pm 0.033$ (Fig. 4e). Due to the limited output power at a pump power of 66.4 μW, the presented data was collected during a period of 20 minutes and at a count rate of 20 KHz. Even for this measurement, $g^2(0)$ is still considerably smaller than 2- suggesting that the transition from chaotic to coherent is quite gradual. All the measurements are fitted in the same manner as the ASE data, maintaining the previously determined coherence time of 3.75 ns. Further investigation of the characteristics of the emission at yet lower pump powers could not be carried out with our current set-up due to the limited output power near the threshold and the low efficiency of the detectors.

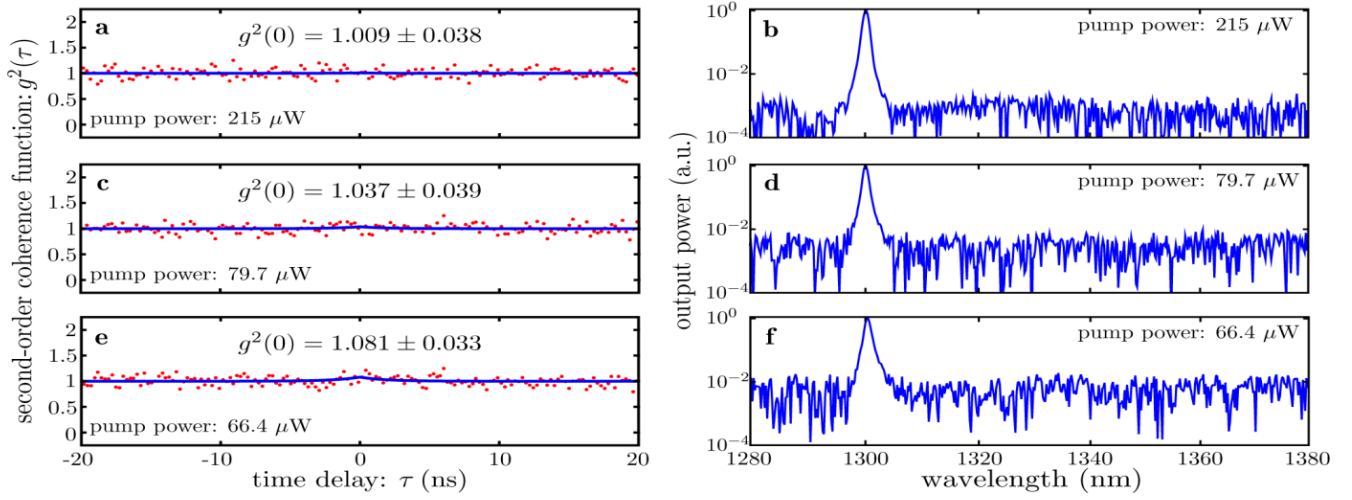

**Fig. 4.** Second-order coherence measurement results for nanoscale coaxial laser in Section 2. The $g^2(\tau)$ measurements of the emitted light from the coaxial nanolaser as well as the corresponding emission spectra in a semi-logarithmic scale at pump powers of **a, b,** 215 µW, **c, d,** 79.7 µW, and **e, f,** 66.4 µW.

The second-order coherence function of the nanoscale coaxial laser reported in Fig. 4 matches theoretical predictions for lasers with high spontaneous emission coupling factor ($\beta$), where just below threshold a small but yet notable component of spontaneous emission is present, while at and well above threshold this contribution diminishes and the emission approaches that of an ideal coherent source [18]. Surprisingly, even around threshold, the radiation from this device appears to be quite coherent.

We also measured the second-order coherence function for a number of disk-shaped nanocavities with various radii. These cavities share an almost identical structure to the coaxial nanolaser with the exception that the silver core is replaced with the gain material. Figure 5 displays the second-order coherence functions along with logarithmic scale single-shots of the emission spectra for two of the example disk-shaped resonators (with radii of 250 nm and 900 nm). The 250 nm radius disk with a single mode emission is studied above threshold. The resulting intensity coincidence peak is $g^2(0) = 1.022 \pm 0.038$, confirming that this light emitting device can generate coherent radiation (see Figs. 5a-b). The electromagnetic simulations for this cavity suggest that the spontaneous emission coupling factor is on the order of $\beta = 0.22$. Next, the larger disk with a radius of 900 nm is investigated (see Figs. 5c-d). For this device, the measured $g^2(0)$ is $1.485 \pm 0.043$. As it is clear in Fig. 5d, the disk resonator with a radius of a 900 nm supports several competing modes. It seems that the simultaneous presence of multiple modes can cause the emission to become more chaotic [31]. Whether the observed $g^2(0) > 1$ is an indication that this device is operating below threshold or the appearing peak at zero time delay is caused due to the beating between independent modes is yet to be fully investigated. Such thorough investigations require detectors with higher timing resolutions. It should be noted that the deviation from $g^2(0) = 1$ in larger nanocavities was regularly observed when we characterized nanolasers with multi-moded spectra. This trend clearly departs from the intensity coincidence measurement reported in [12], where a device supporting two modes of nearly equal amplitudes generate a $g^2(0) = 1$.

## 5. CONCLUSIONS

In conclusion, the areas of nanophotonics and plasmonics have progressed tremendously in the past couple of decades. Undoubtedly, the introduction of metallic structures has opened a path towards light confinement and manipulation at the subwavelength scale – a regime that was previously thought to be out of reach in optics. Of central importance in this endeavor is to devise subwavelength light emitting devices that can power up the future nanoscale photonic circuits. The metal-clad coaxial and disk-shaped nanoresonators can provide viable platforms to implement such subwavelength sources. They support ultra-small cavity modes and offer large mode-emitter overlap as well as multifold scalability. In addition, because of their small size and high Purcell factor, metallic nanolasers are expected to show large direct modulation bandwidths [32]. Furthermore, coherent radiation generally has a lower relative intensity noise (RIN) in comparison to incoherent light from LEDs, hence metallic nanoscale *lasers* are expected to operate more reliably as high-speed devices [33].

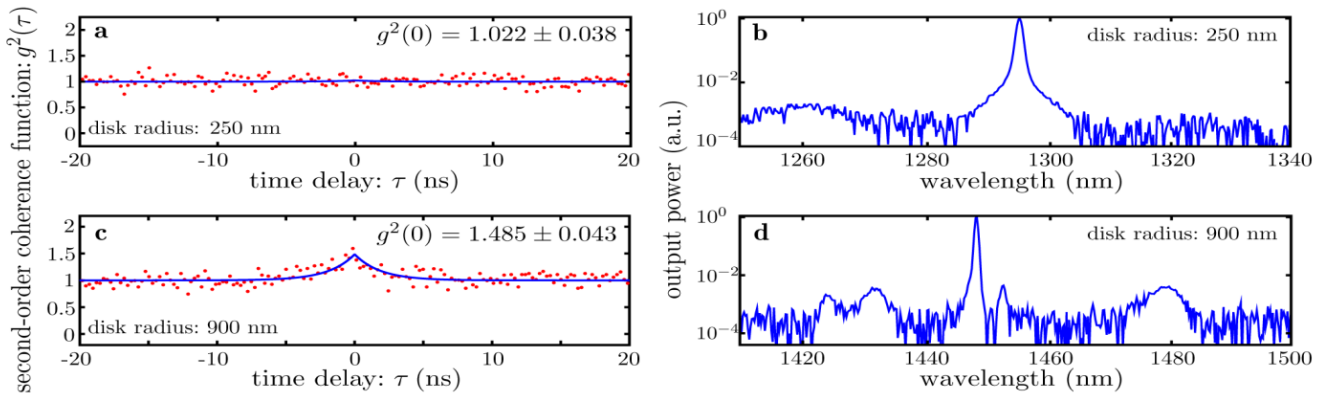

**Fig. 5.** Second-order coherence measurement results for nanoscale disk-shape lasers. The $g^2(\tau)$ measurements of the emitted light from the disk-shaped nanolaser as well as the corresponding emission spectra in a semi-logarithmic scale for disk radii of **a, b,** 250 nm, **c, d,** 900 nm.

In this manuscript, we reported our measurement results for the second-order coherence functions of coaxial and disk-shaped nanoscale lasers (with InGaAsP multiple quantum well gain systems). These measurements were accomplished by establishing a set-up to reliably characterize the intensity correlation function for broad linewidth sources at telecommunication bands. In order to ensure the capability of our setup to measure $g^2(\tau)$, the second order coherence is first measured for a spectrally broad (~50 nm) commercial ASE source. In addition, by optimizing the design and modifying the fabrication process, a number of nanolasers developed are capable of generating *relatively high output power* (few to few tens of microwatts) and could operate under CW pumping for an extended time.

The second-order coherence measurement results presented in Section 4 of this manuscript unambiguously confirm that nanoscale coaxial and disk-shaped metallic cavities can indeed generate coherent radiation ($g^2(0) \sim 1$). Further investigations of second-order coherence properties of nanolasers below threshold and for devices with higher spontaneous emission coupling factors require single photon counters with yet higher timing resolutions and efficiencies. These studies may in turn shed light on the quantum properties of the emission from metallic nanoscale light sources.

**Funding.** Army Research Office (ARO) (W911NF-16-1-0013, W911NF-14-1-0543).

**Acknowledgment.** The authors would like to thank Bruno Sanguinetti from ID Quantique, Demetrios Christodoulides, Parinaz Aleahmad and Nicholas Nye from CREOL for fruitful discussions. The authors also appreciate ID Quantique for generously loaning the single photon detectors.

See Supplementary 1 for supporting content.